\documentclass[9pt,letterpaper,aps,pra,notitlepage,twocolumn,reprint,shortbibliography,superscriptaddress,nofootinbib]{revtex4-1}
\usepackage[utf8]{inputenc}
\usepackage{amsmath,amssymb}
\usepackage{graphicx}
\usepackage{hyperref}
\usepackage{tikz}
\usepackage{diagbox}
\usepackage[margin=2cm]{geometry}
\usepackage{bm}
\newcommand{\ket}[1]{| #1 \rangle}

\newcommand{\eq}[1]{\begin{align}#1\end{align}}

\newcommand{\pr}[1]{| #1 \rangle \langle #1|}

\newcommand{\vp}{\bm{T}}
\newcommand{\vn}{\bm{W}}
\usepackage{color}
\usepackage{bbm}
\usepackage{blindtext}
\usepackage{comment}

\newcommand{\beq}{\begin{equation}}
\newcommand{\eeq}{\end{equation}}

\definecolor{JM}{RGB}{4,116,149}

\definecolor{nico}{RGB}{51, 87, 255}

\usepackage{pgfplots}

\begin{document}
	\title{Quadratic speedup for simulating Gaussian boson sampling}
	\author{Nicol\'as Quesada}\thanks{Equal contributors}
	
	\affiliation{Xanadu, 777 Bay Street, Toronto, Canada}
	\author{Rachel S. Chadwick}\thanks{Equal contributors}
	\affiliation{Quantum Engineering Technology Labs, University of Bristol, Bristol BS8 1FD, UK}
	\affiliation{Quantum Engineering Centre for Doctoral Training, University of Bristol, Bristol BS8 1FD, UK}
	\author{Bryn A.  Bell}\thanks{Equal contributors}
	\affiliation{Ultrafast Quantum Optics group, Department of Physics,Imperial College London, London SW7 2AZ, UK}
	\author{\\ Juan Miguel Arrazola}
	\affiliation{Xanadu, 777 Bay Street, Toronto, Canada}
	\author{Trevor Vincent}
	\affiliation{Xanadu, 777 Bay Street, Toronto, Canada}	

	\author{Haoyu Qi}
	\affiliation{Xanadu, 777 Bay Street, Toronto, Canada}	
	\author{Ra\'ul Garc\'ia-Patr\'on}
	\affiliation{School of Informatics, University of Edinburgh, Edinburgh EH8 9AB, UK}
		
\begin{abstract}
We introduce an algorithm for the classical simulation of Gaussian boson sampling that is quadratically faster than previously known methods. The complexity of the algorithm is exponential in the number of photon pairs detected, not the number of photons, and is directly proportional to the time required to calculate a probability amplitude for a pure Gaussian state. The main innovation is to use auxiliary conditioning variables 
to reduce the problem of sampling to computing pure-state probability amplitudes, for which the most computationally-expensive step is calculating a loop hafnian. We implement and benchmark an improved loop hafnian algorithm and show that it can be used to compute pure-state probabilities, the dominant step in the sampling algorithm, of up to 50-photon events in a single workstation, i.e., without the need of a supercomputer.
\end{abstract}

\maketitle
\section{Introduction}
Building quantum computers capable of convincingly performing tasks that are intractable to replicate using classical computers is a significant technological milestone~\cite{arute2019quantum, harrow2017quantum}. For photonic implementations, boson sampling and its variants are a leading approach for demonstrating such capabilities~\cite{aaronson2011computational, barkhofen2017driven,lund2014boson,lund2017exact,chakhmakhchyan2017boson, bentivegna2015experimental, wang2019boson,deshpande2021quantum}. Gaussian Boson Sampling (GBS) has emerged as a platform to tackle challenges in scaling to larger numbers of photons and modes~\cite{hamilton2017gaussian,kruse2018detailed}. GBS has already been experimentally implemented~\cite{clements2018approximating, paesani2019generation, zhong2019experimental,zhong2020quantum,zhong2021phase,arrazola2021quantum}, and it has generated additional interest due to the discovery of algorithms and applications~\cite{huh2015boson,arrazola2018using, arrazola2018quantum, banchi2019molecular, schuld2019quantum,jahangiri2019point, banchi2020training, jahangiri2020quantum}. These advances make GBS a leading candidate for demonstrating a quantum advantage in photonics.

In addition to experimental implementations significant progress has also been made in developing classical simulation algorithms. 
For boson sampling, a first method was an approximate Markov chain Monte Carlo algorithm~\cite{neville2017classical}. This was improved in Ref.~\cite{clifford2018classical}, where an exact sampling algorithm was introduced such that the complexity for generating one sample with $N$ photons is equivalent to that of calculating an output probability amplitude. This in turn is equivalent to computing the permanent of an $N\times N$ matrix, which requires $O(N \, 2^N)$ time using the best known methods~\cite{ryser1963combinatorial}.

An effort to obtain simulation techniques for GBS has also been pursued. An exact algorithm has been reported and implemented for GBS with threshold detectors~ \cite{quesada2018gaussian,gupt2018classical}, but it suffers from exponential memory requirements. Two algorithms were also proposed in Ref.~\cite{wu2020speedup} for a restricted version of GBS. The first one has polynomial space complexity and $O(\text{poly}(N)2^{8N/3} )$ time complexity; the second has exponential space complexity and $O(\text{poly}(N)2^{5 N/2})$ time complexity. Recently, an exact sampling algorithm for GBS was presented that requires only polynomial memory~\cite{quesada2020exact}. This algorithm shows an improved complexity proportional to $O(N^3 2^N)$ for generating one sample with $N$ photons. These GBS algorithms mark a crucial difference with respect to boson sampling: in GBS with pure states, the probability of observing an outcome with $N$ photons is proportional to the loop hafnian of an $N\times N$ matrix, for which the best algorithms require $O(N^3 2^{N/2})$ time. In other words, there is a quadratic gap between the complexity to generate a sample and the complexity to compute an output probability. This suggests the possible existence of a better sampling algorithm with complexity matching that of calculating an output probability amplitude, similar to what has been achieved for boson sampling.

We present such an algorithm. The main insight is to introduce conditioning auxiliary variables obtained by performing a virtual heterodyne measurement in all modes and iteratively replace the heterodyne outcomes with photon number measurements. The photon number outcomes conditional on the heterodyne measurements are given by pure-state probabilities. We develop a chain of conditional probabilities, where at each step only probabilities of a pure Gaussian state need to be calculated. For outputs with $N$ photons, these are proportional to loop hafnians of $N\times N$ matrices, which can be calculated in $O(N^3  2^{N/2})$ time \cite{bjorklund2018faster}. In general, the number of output probabilities that must be calculated is proportional to the number of modes $m$, which leads to a time complexity upper bounded by $O(m N^3  2^{N/2})$. This corresponds to a quadratic speedup over the previous state-of-the-art. It also suggests that, compared to boson sampling, roughly twice as many photons are needed in GBS to reach the regime where classical simulations become intractable. We implement an improved version of a loop hafnian algorithm and use it to compute pure-state probabilities of events with up to 50 photons using a workstation with 96 CPUs.

In what follows, we begin by giving a short overview of GBS in Sec.~\ref{sec: GBS}. We then describe our simulation algorithm in Sec. \ref{sec: algorithm}, which constitutes our main result. In Sec.~\ref{sec:benchmark}, we benchmark an algorithm for computing loop hafnians, which determine the most expensive step for simulation, and finally present a discussion of our findings in Sec.~\ref{sec:discussion}.

\section{Gaussian boson sampling}\label{sec: GBS}

The quantum state of a collection of $m$ optical modes can be uniquely described in terms of its Wigner function~\cite{weedbrook2012gaussian,serafini2017quantum}. Gaussian states are states whose Wigner function is a Gaussian distribution. They can be described by a covariance matrix $\bm{V}$ and a vector of means $\bar{\bm{R}} =\left[ \begin{smallmatrix} \bm{\bar{q}} \\ \bm{\bar{p}} \end{smallmatrix} \right]$, where $\bm{\bar{q}},\ \bm{\bar{p}} $ are the mean canonical position and momentum vectors. It is also possible to express the covariance matrix in terms of the complex amplitudes $\bm{\alpha} = \tfrac{1}{\sqrt{2 \hbar }} (\bm{q}+i \bm{p}) \in \mathbb{C}^m$, which are described by a complex-normal distribution with mean $\bm{\bar{\alpha}} = \tfrac{1}{\sqrt{2 \hbar}} (\bm{\bar{q}}+i \bm{\bar{p}}) \in \mathbb{C}^m$ and covariance matrix~\cite{picinbono1996second},
\eq{
\bm{\Sigma}	= \bm{F} \bm{V} \bm{F}^\dagger, \quad \bm{F} := \frac{1}{\sqrt{2 \hbar}} \begin{bmatrix} \mathbbm{I} & i \mathbbm{I} \\ \mathbbm{I} & -i \mathbbm{I} \end{bmatrix}.
}
where $\mathbbm{I}$ is the identity matrix.

Gaussian Boson Sampling (GBS) is a form of photonic quantum computing where a Gaussian state is measured in the photon-number basis.
If $\bm{\bar{\alpha}} = 0$ , the probability of observing the output sample $S = (s_m,\ldots,s_1)$, where $s_i$ is the number of photons in mode $i$, is given by~\cite{hamilton2017gaussian}

\begin{align}\label{eq:GBS}
p(S) = \frac{1}{\sqrt{\text{det}(\bm{Q})} } \frac{\text{haf}(\bm{A}_S)}{s_1!\cdots s_m!}, 
\end{align}
where 
\begin{align}
\bm{Q}&:=\bm{\Sigma} +\mathbbm{I}/2,\\
\bm{A} &:= \bm{X} \left(\mathbbm{I} - \bm{Q}^{-1}\right),\label{Eq:A_matrix}\\
\bm{X} &:=  \left[\begin{smallmatrix}
	0 &  \mathbbm{I} \\
	\mathbbm{I} & 0  
\end{smallmatrix} \right],
\end{align}
and $\bm{A}_S$ is the matrix obtained as follows: if $s_i=0$, rows and columns $i$ and $i+m$ are deleted from $\bm{A}$; if $s_i>0$, the rows and columns are repeated $s_i$ times. The hafnian of a square symmetric matrix of even dimension $n$ is defined as
\eq{\label{eq:defhaf}
\text{haf}(\bm{C}) = \sum_{X \in \text{PMP}(n)} \prod_{(i,j) \in X} \bm{C}_{i,j},
}
where $\text{PMP}(n)$ is the set of perfect matching permutations~\cite{bjorklund2018faster}.

Defining $N:=\sum_i s_i$ as the total number of photons, the submatrix $\bm{A}_S$ has dimension $2N$, meaning that the best algorithm for calculating its hafnian requires $O(N^3 2^{2N/2})=O(N^3 2^N)$ time. However, when the Gaussian state is pure, it is possible to write $\bm{A}=\bm{B}\oplus \bm{B}^*$ and the probability distribution reduces to~\cite{hamilton2017gaussian}
\begin{align}\label{eq:GBS-pure}
p(S) = \frac{1}{\sqrt{\text{det}(\bm{Q})} } \frac{|\text{haf}(\bm{B}_S)|^2}{s_1!\cdots s_m!}, 
\end{align}
where $\bm{B}_S$ is constructed analogously: if $s_i=0$, rows and columns $i$ are deleted; if $s_i>0$, the rows and columns are repeated $s_i$ times. Since the matrix $\bm{B}_S$ has dimension $N$, computing its hafnian requires only 
$O(N^3 2^{N/2})$ time. 

An analogous formula can also be derived for GBS with displacements, i.e., when $\bm{\bar{\alpha}} \neq 0$. We define the quantities:
\begin{align}
\vec \alpha &= \begin{bmatrix}\bm{\bar \alpha} \\ \bm{\bar \alpha} ^* \end{bmatrix} = \bm{F} \bar{\bm{R}}, \\
\vec \gamma  &= \begin{bmatrix} \bm{\bar \gamma} \\  \bm{\bar \gamma}^*\end{bmatrix} = 
\bm{Q}^{-1} \vec \alpha ,\\
p(\text{vac}) &=\frac{\exp\left(-\tfrac{1}{2} \left[ \vec{\alpha} \right]^\dagger  \left[ \bm{Q} \right]^{-1} \vec{\alpha} \right) }{ \sqrt{\text{det}(\bm{Q})} }.
\end{align}
In this case, the output probabilities are given by~\cite{bjorklund2018faster,quesada2019simulating}:
\begin{align}\label{Eq: lhaf}
p(S)  = p(\text{vac})  \frac{\text{lhaf}\left\{\text{filldiag} \left( {\bm{A}}_S , \vec{\gamma}_S \right)\right\}}{s_1!\cdots s_m!},
\end{align}
where $\text{lhaf}(\cdot)$ is the \emph{loop hafnian}~\cite{bjorklund2018faster}, $\vec \gamma_S$ is obtained from $\vec \gamma$ by repeating the $i, i+m$  entries of $\vec \gamma$ a total of $s_i$ times, and $\text{filldiag} \left( {\bm{A}}_S , \vec{\gamma}_S \right)$ replaces the diagonal of ${\bm{A}}_S$ with the vector $\vec{\gamma}_S$. The loop hafnian of an arbitrary matrix is defined analogously to the hafnian in Eq.~\eqref{eq:defhaf}, but replacing the set $\text{PMP}(n)$ by the set of single-pair matchings $\text{SPM}(n)$~\cite{bjorklund2018faster}.

When the Gaussian state is pure, it is possible to express the output probability as 
\begin{align}\label{Eq: lhaf-pure}
p(S)  =    p(\text{vac})\frac{ \left| \text{lhaf}\left\{\text{filldiag} \left( {\bm{B}}_S , \bar{\bm{\gamma}}_S \right)\right\}\right|^2}{s_1!\cdots s_m!},
\end{align}
where $\bar{\bm{\gamma}}_S$ is obtained from $\bar{\bm \gamma}$ by repeating its $i$-th entry $s_i$ times.

Up to constant prefactors, the best known algorithms for calculating loop hafnians of generic matrices have the same complexity as for hafnians, namely $O(N^3 2^{N/2})$ for matrices of dimension $N$. This implies that the complexity of computing output probabilities with $N$ photons for pure Gaussian states with displacements also scales as $O(N^3 2^{N/2})$. We make use of this result in the next section.

So far we have focused only on describing the photon-number statistics of a multimode Gaussian state. For the sampling algorithm we describe in the next section it will also be necessary to recall some basic properties of continuous-output measurements on a Gaussian state.  Of particular relevance here are so-called heterodyne measurements.  The outcomes of an $m$-mode heterodyne measurement are specified by a vector of complex numbers $\bm{\alpha} = \tfrac{1}{\sqrt{2\hbar}} \left(\bm{q}+ i \bm{p} \right) \in \mathbb{C}^m$.
Generating heterodyne samples from a Gaussian state is straightforward. 
For a Gaussian state with vector of means $\bar{\bm{R}}$ and covariance matrix $\bm{V}$, we sample from the multivariate normal distribution $\bm{\mu} \sim \mathcal{N}(\bar{\bm{R}},\,\bm{V}_Q)$, where $\bm{\mu}=\left[ \begin{smallmatrix}\bm{q}\\ \bm{p}\end{smallmatrix}\right]$ and $\bm{V}_Q$ is the Husimi $Q$-function covariance matrix \cite{husimi1940some} in the quadrature basis, given by
\begin{equation}
    \bm{V}_Q=\bm{V}+\frac{\hbar}{2}\mathbbm{I}.
\end{equation}
Explicitly, we obtain outcome $\bm{\mu}$ with probability 
\begin{equation}
    p( \bm{\mu})=  \frac{\exp\left( -\tfrac{1}{2} (\bm{\mu} - \bar{\bm{R}})^T  \ \bm{V}_Q^{-1} \ (\bm{\mu} - \bar{\bm{R}}) \right)}{\sqrt{\det(2 \pi  \bm{V}_Q)}}
    \label{eq:heterodyne}.
\end{equation}
A partial heterodyne measurement can be performed when measuring only a subset of the modes. This can be done by either sampling from the reduced matrix of $\bm{V}_Q$, formed by selecting only the rows and columns of the included modes, or by sampling all modes and discarding the outcomes for modes we don't wish to sample.

It is also useful to write the conditional state of a subset of the modes, $\mathcal{A}$, when the other modes, $\mathcal{B}$, are measured using heterodyne detectors. 
For this it is convenient to write the covariance matrix $\bm{V}$ in block form with modes in $\mathcal{A}$ and $\mathcal{B}$ in separate blocks, and similarly to group the modes in the vector of means. This can be done by permuting the rows and columns in the covariance matrix and the elements of the vector of means, keeping the ordering in both consistent. So we write both the covariance matrix and vector of means in the ordering $(\bm{q}_\mathcal{A},\bm{p}_{\mathcal{A}}, \bm{q}_{\mathcal{B}}, \bm{p}_{\mathcal{B}})$:
\begin{equation}
    \bm{V}=\left(
    \begin{array}{cc}
    \bm{V}_{\mathcal{A}\mathcal{A}}     &  \bm{V}_{\mathcal{A}\mathcal{B}}\\
    \bm{V}_{\mathcal{B}\mathcal{A}}     & \bm{V}_{\mathcal{B}\mathcal{B}}
    \end{array}
    \right),
    \hspace{1cm}
    \bar{\bm{R}}=\left[ \begin{array}{c}\bar{\bm{R}}_\mathcal{A} \\\bar{\bm{R}}_\mathcal{B} \end{array}
    \right].
\end{equation}
If the outcome $\bm{\mu}_\mathcal{B}$ is obtained by measuring the modes in the set $\mathcal{B}$, we can write the resulting covariance matrix and vector of means for $\mathcal{A}$ as \cite{serafini2017quantum}
\begin{align}
    \bm{V}_{\mathcal{A}}^{(\mathcal{B})}&=\bm{V}_{\mathcal{A}\mathcal{A}}-\bm{V}_{\mathcal{A}\mathcal{B}}(\bm{V}_{\mathcal{B}\mathcal{B}}+\tfrac{\hbar}{2}\mathbbm{I})^{-1}\bm{V}_{\mathcal{B}\mathcal{A}}, \label{eq:condV}\\
    \bm{R}_\mathcal{A}^{(\mathcal{B})}&=\bm{\bar{R}}_\mathcal{A}+\bm{V}_{\mathcal{A}\mathcal{B}}(\bm{V}_{\mathcal{B}\mathcal{B}}+\tfrac{\hbar}{2}\mathbbm{I})^{-1}(\bm{\mu}_\mathcal{B}-\bm{\bar{R}}_\mathcal{B}). \label{eq:condR}
\end{align}
This update rule together with the sampling in Eq.~\eqref{eq:heterodyne} will be useful in the next section where we introduce our algorithm.

\section{Algorithm} \label{sec: algorithm}

We now describe an algorithm which samples the modes sequentially, but that, unlike the one from Ref. \cite{quesada2020exact}, never requires the calculation of mixed-state probabilities. In this algorithm we introduce partial heterodyne measurements so that only pure-state probabilities need to be evaluated at each step. 
Given a partial heterodyne measurement on modes in $\mathcal{B}$, the
 photon number probability of a pattern $S$ in modes $\mathcal{A}$ is given by Eq.~\eqref{Eq: lhaf-pure}, where $\bm{B}$ and $\bm{\gamma}$ are now found for the state with covariance matrix $\bm{V}_{\mathcal{A}}^{(\mathcal{B})}$ and vector of means $\bm{R}_\mathcal{A}^{(\mathcal{B})}$ as given in Eq.~\eqref{eq:condR}. Note that if the global covariance matrix $\bm{V}$ corresponds to a pure state, the conditional covariance matrix $\bm{V}_{\mathcal{A}}^{(\mathcal{B})}$ also corresponds to a pure state. 

If we wish to sample in the photon-number basis in modes $\mathcal{A}$ without measuring the other modes, we can make use of the observation in the paragraph above: the marginal photon-number probabilities  can be obtained by integrating the joint probabilities over the set of possible heterodyne outcomes $\bm{\alpha}_\mathcal{B}$:
\begin{align}
    p(\bm{s}_\mathcal{A}) &=\int d \bm{\alpha}_\mathcal{B} \, p(\bm{s}_\mathcal{A}, \bm{\alpha}_\mathcal{B}) =\int d \bm{\alpha}_\mathcal{B} \, p(\bm{\alpha}_\mathcal{B}) p(\bm{s}_\mathcal{A}|\bm{\alpha}_\mathcal{B}).
\end{align}
Hence we can sample from the marginal probabilities by sampling from $p(\bm{\alpha}_\mathcal{B})$, followed by the conditional probabilities $p(\bm{s}_\mathcal{A}|\bm{\alpha}_\mathcal{B})$ and then simply ignoring the heterodyne outcome. 
This is justified because one can always sample from a marginal distribution by considering additional virtual variables and then sampling from the correspondingly enlarged probability distribution, as long as one then forgets the values of the added variables. The outcomes of the heterodyne measurements are precisely these added variables; they do not correspond to real measurements in an experimental setup but are rather virtual measurements to introduce convenient conditioning auxiliary variables.

We now want to apply this directly to sampling the modes sequentially. The objective is to sample mode $k$ given the previous modes $1,\ldots,k-1$ have already been sampled, so we set $\mathcal{A}=1,\ldots,k$ and $\mathcal{B}=k+1,\ldots,m$. We assume that we have already sampled from $p(s_1,\ldots,s_{k-1},\alpha_{k+1},\ldots,\alpha_m)$. We wish to sample $s_k$ conditional on this outcome. To do this we can calculate the relative probabilities for all $s_k$ and sample from that distribution. This will result in a sample drawn from $p(s_1,\ldots,s_k,\alpha_{k+1},\ldots,\alpha_m)$. By discarding $\alpha_{k+1}$, we are left with a sample from $p(s_1,\ldots,s_k,\alpha_{k+2},\ldots,\alpha_m)$ and are ready to sample the $(k+1)^\text{th}$ mode. We work progressively from $k=1$ to $m$ essentially replacing the virtual heterodyne measurements with photon-number measurements until we are left with a sample from $p(s_1,\ldots,s_m)$.

This algorithm can sample from a pure multimode state, yet in realistic experiments, the full state may not be pure, for example if loss is included. To address this, we express the mixed Gaussian state as a convex combination of pure states. The Williamson decomposition~\cite{williamson1936algebraic,weedbrook2012gaussian} of a quantum covariance matrix $\bm{V}$ states that it can be split as
\eq{\label{eq:split}
	\bm{V} = \vp+\vn,
}
where $\vp = \tfrac{\hbar}{2}\bm{S} \bm{S}^T$ is the covariance matrix of a pure state, $\bm{S}$ is a symplectic matrix, and $\vn$ is positive semidefinite. In Hilbert space, this implies that a mixed Gaussian state with covariance matrix $\bm{V} = \vp+\vn$ and a vector of means $\bar{\bm{R}} = \left[\begin{smallmatrix}\bar{\bm{q}}\\ \bar{\bm{p}}\end{smallmatrix}\right]$ can be expressed as~\cite{wolf2004gaussian, serafini2017quantum}
\beq\label{eq:decomp} 
\varrho = \int d \bm{R} \  	p( \bm{R}) \ 
\pr{\psi_{\bm{R}, \vp}},
\eeq
where  
\eq{\label{eq:mnpdf}
		p( \bm{R})&=  \frac{\exp\left( -\tfrac{1}{2} (\bm{R} - \bar{\bm{R}})^T  \ \vn^{-1} \ (\bm{R} - \bar{\bm{R}}) \right)}{\sqrt{\det(2 \pi  \vn)}},
} 
is the probability density function of a multivariate normal distribution, and $\ket{\psi_{\bm{R}, \vp}}$ is a pure Gaussian state with vector of means $\bm{R}$ and covariance matrix $\vp$. Using this, the probability of observing an outcome $(s_1, s_2, \ldots, s_m)$ when performing a measurement on a mixed state $\varrho$ on $m$ modes is
\eq{
p(s_1, \ldots, s_m)&=\langle s_1, \ldots, s_m| \varrho | s_1, \ldots, s_m\rangle\nonumber\\
&=\int d \bm{R} \  	p( \bm{R}) |\langle\psi_{\bm{R}, \vp}|s_1, \ldots, s_m\rangle|^2\nonumber\\
&=\int d \bm{R} \  	p( \bm{R}) p(s_1, \ldots, s_m|\bm{R}).
}
Hence to sample from a mixed state we can sample the displacement vector from Eq.~\eqref{eq:mnpdf} and then sample from the resulting pure state.

We can now outline the full algorithm for simulating Gaussian boson sampling:\\

\textbf{Algorithm:}
To sample from a state given by covariance matrix $\bm{V}$ and vector of means $\bm{\bar{R}}$ in $m$ modes:
\begin{enumerate}
	\item If the Gaussian state to be sampled is mixed, calculate the matrices $\vp,\,\vn$ from the Williamson decomposition such that $\bm{V} = \vp+\vn$. Sample a vector $\bm{R}$ from the multivariate normal distribution $p( \bm{R})$ as in Eq.~\eqref{eq:mnpdf}. This can be done in cubic time in the size of the matrix~\cite{gentle2009computational}. Continue the algorithm with the pure state given by covariance matrix $\vp$ and vector of means $\bm{R}$. 
	\item Generate a sample from the probability distribution 
    $p(\alpha_2,\ldots,\alpha_m)$ resulting from heterodyne measurements on modes 2 to $m$ using Eq.~\eqref{eq:heterodyne}. 
    This can be done in cubic time in the number of modes.
\end{enumerate}
For modes $k=1,\ldots,m$:
\begin{enumerate}
    \setcounter{enumi}{2}
    \item Given the sample from $p(s_1,\ldots,s_{k-1},\alpha_{k+1},\ldots,\alpha_m)$ (which is found in the previous step), find $\bm{V}^{(k)}$ and $\bm{R}^{(k)}$ the covariance matrix and vector of means for the first $k$ modes conditional on the measurement $(\alpha_{k+1},\ldots,\alpha_m)$ in the other modes, using Eqs.~\eqref{eq:condV} and ~\eqref{eq:condR}. For $k=1$ we start with a sample from $p(\alpha_2,\ldots, \alpha_m)$.
    
	\item Given a cutoff $d$, use Eq.~\eqref{Eq: lhaf-pure} with $\bm{V}^{(k)}$ and $\bm{R}^{(k)}$ to compute the probabilities $p(s_k,s_{k-1}^\star,\ldots,s_1^\star|\alpha_{k+1}^\star,\ldots,\alpha_m^\star)$ for $s_k=0,1,\ldots, d$, with all starred variables sampled previously from the correct distribution. This is the most computationally-demanding step, which requires computing loop hafnians. 
	\item Sample $s_k$ with respect to the distribution
	\eq{
	\frac{p(s_k,s_{k-1}^\star,\ldots, s_{1}^\star|\alpha_{k+1}^\star,\ldots,\alpha_m^\star)}{\sum_{s_k}p(s_k,s_{k-1}^\star,\ldots, s_{1}^\star|\alpha_{k+1}^\star,\ldots,\alpha_m^\star)}.
	}
	\item Discard $\alpha_{k+1}^\star$ to get an outcome sampled from $p(s_1,\ldots,s_k,\alpha_{k+2},\ldots,\alpha_m)$.
\end{enumerate}
After repeating steps 3-6 for each $k=1,2,\ldots,m$, we are left with an outcome sampled from $p(s_1,\ldots,s_m)$.
The correctness of the algorithm follows directly from the definition of conditional probabilities and integration over the auxiliary variables $\alpha_2,\ldots,\alpha_m$ as shown in Appendix \ref{app:correctness}.

Overall, the algorithm reduces sampling from the GBS distribution to computing pure-state probabilities $p(s_k,s_{k-1}^\star,\ldots, s_{1}^\star|\alpha_{k+1},\ldots,\alpha_m)$. When a total of $N$ photons are detected, calculating the largest such probability amplitude requires $O(N^3 2^{N/2})$ time, which results from computing loop hafnians in Step 4. This is scaled up by at most the total number of modes, giving a total sampling complexity of $O(m N^3 2^{N/2})$. This is a quadratic improvement over the algorithm in Ref.~\cite{quesada2020exact} which has complexity $O(m N^3 2^{N})$.
 
It is worthwhile to compare to the algorithm of Ref.~\cite{clifford2018classical} for boson sampling, whose complexity is $O(N 2^{N})$. Up to polynomial factors, our algorithm suggests that the time required to simulate boson sampling with $N$ photons is roughly the same as simulating GBS with $2N$ photons.
We note that our algorithm can also be used to simulate GBS with threshold detectors: once a photon number sample is generated, simply set $s_i=1$ if $s_i>0$, where $s_i=1$ denotes that the detector measured one or more photons.

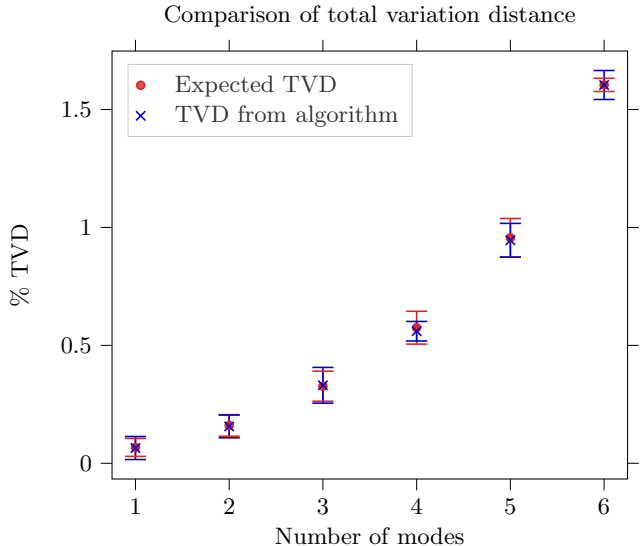
\begin{figure}[t!]
		\centering
\begin{tikzpicture}

\definecolor{color0}{rgb}{0.83921568627451,0.152941176470588,0.156862745098039}

\begin{axis}[
legend cell align={left},
legend style={fill opacity=0.8, draw opacity=1, text opacity=1, at={(0.03,0.97)}, anchor=north west, draw=white!80!black},
tick align=outside,
tick pos=both,
title={Comparison of total variation distance},
x grid style={white!69.0196078431373!black},
xlabel={Number of modes},
xmin=0.75, xmax=6.25,
xtick style={color=black},
y grid style={white!69.0196078431373!black},
ylabel={\% TVD},
ymin=-0.0666038617570655, ymax=1.74795658814131,
ytick style={color=black}
]
\path [draw=color0, semithick]
(axis cs:1,0.0293519652952263)
--(axis cs:1,0.105347082114494);

\path [draw=color0, semithick]
(axis cs:2,0.11471713394611)
--(axis cs:2,0.206199927598768);

\path [draw=color0, semithick]
(axis cs:3,0.263047980576645)
--(axis cs:3,0.390566403563393);

\path [draw=color0, semithick]
(axis cs:4,0.505292406010212)
--(axis cs:4,0.644524814207016);

\path [draw=color0, semithick]
(axis cs:5,0.874553896756856)
--(axis cs:5,1.03797357270887);

\path [draw=color0, semithick]
(axis cs:6,1.57649255756795)
--(axis cs:6,1.63247907693181);

\path [draw=blue!80.3921568627451!black, semithick]
(axis cs:1,0.0158761586928607)
--(axis cs:1,0.113628986088322);

\path [draw=blue!80.3921568627451!black, semithick]
(axis cs:2,0.107942242190148)
--(axis cs:2,0.204564545004983);

\path [draw=blue!80.3921568627451!black, semithick]
(axis cs:3,0.254903482603554)
--(axis cs:3,0.406547523040498);

\path [draw=blue!80.3921568627451!black, semithick]
(axis cs:4,0.518694189536256)
--(axis cs:4,0.60148341692745);

\path [draw=blue!80.3921568627451!black, semithick]
(axis cs:5,0.874126613235152)
--(axis cs:5,1.0172386181603);

\path [draw=blue!80.3921568627451!black, semithick]
(axis cs:6,1.54259748412832)
--(axis cs:6,1.66547656769138);

\addplot [semithick, color0, mark=*, mark size=1.5, mark options={solid}, only marks]
table {%
1 0.0673495237048602
2 0.160458530772439
3 0.326807192070019
4 0.574908610108614
5 0.956263734732863
6 1.60448581724988
};
\addlegendentry{\ Expected TVD}
\addplot [semithick, blue!80.3921568627451!black, mark=x, mark size=2.5, mark options={solid}, only marks]
table {%
1 0.0647525723905912
2 0.156253393597565
3 0.330725502822026
4 0.560088803231853
5 0.945682615697726
6 1.60403702590985
};
\addlegendentry{\ TVD from algorithm}
\addplot [semithick, color0, mark=-, mark size=4, mark options={solid}, only marks]
table {%
1 0.0293519652952263
2 0.11471713394611
3 0.263047980576645
4 0.505292406010212
5 0.874553896756856
6 1.57649255756795
};
\addplot [semithick, color0, mark=-, mark size=4, mark options={solid}, only marks]
table {%
1 0.105347082114494
2 0.206199927598768
3 0.390566403563393
4 0.644524814207016
5 1.03797357270887
6 1.63247907693181
};
\addplot [semithick, blue!80.3921568627451!black, mark=-, mark size=4, mark options={solid}, only marks]
table {%
1 0.0158761586928607
2 0.107942242190148
3 0.254903482603554
4 0.518694189536256
5 0.874126613235152
6 1.54259748412832
};
\addplot [semithick, blue!80.3921568627451!black, mark=-, mark size=4, mark options={solid}, only marks]
table {%
1 0.113628986088322
2 0.204564545004983
3 0.406547523040498
4 0.60148341692745
5 1.0172386181603
6 1.66547656769138
};
\end{axis}

\end{tikzpicture}
		\caption{The total variation distance (TVD) of the proposed algorithm and the expected TVD computed by estimating probabilities from brute force sampling of the ideal distribution. We numerically find the TVD between the exact probabilities and the estimated probabilities for all outputs below the cutoff photon number. The expected TVD was estimated by sampling from the exact distribution. For both sampling algorithms, the probability distribution was estimated by taking a sample of size 200000, and averaging over ten Haar random unitaries. The squeezing parameter was 0.5 in all modes, giving an average photon number of $\sim 0.27$ per mode, and we chose a cutoff of six photons. The error bars show the standard deviation of the ten repeated tests. } \label{Fig:TVD}
\end{figure}

Finally, note that each random variable $s_k$ has support over the non-negative integers, thus in principle one needs to calculate all the (infinitely many) conditional probabilities. However, we can choose some cutoff number of photons $d$ in each mode such that the probability of getting a photon number above this value is negligible. We confirm the accuracy of the algorithm in Fig. (\ref{Fig:TVD}) where we show that the total variation distance lies within the expected range for the sample size as compared to a brute force sampler with the same number of samples. This shows that the chosen cutoff was sufficiently high to not introduce any notable error.

\section{Benchmarking}\label{sec:benchmark}
In this section, we test the performance of a new implementation of the loop hafnian algorithm of Ref.~\cite{bjorklund2018faster}, which is available in \texttt{The Walrus}~\cite{code}. 
The evaluation of loop hafnians is delegated to multi-threaded C++ code which uses the La Budde algorithm~\cite{rehman2011budde} for calculating the characteristic polynomial of a matrix. This gives a speedup of about three times with respect to previous algorithms based on diagonalization, but more importantly, improves significantly the accuracy of the calculation.

In the original implementation of the loop hafnian algorithm, which uses double precision and eigenvalue methods for calculating power traces, it was found that a relative error of $\sim 10^{-1}$ is present for computing the loop hafnian of the $54 \times 54$ all-ones matrix~\cite{bjorklund2018faster}. To get around this issue, we use the aforementioned La Budde algorithm to significantly improve the accuracy of the calculation of the characteristic polynomial of a matrix~\cite{rehman2011budde}. Moreover, this method allows us to use long double complex data types. With these changes we lower the relative error in the calculation by three orders of magnitude compared to the previous implementation. We can then achieve a precision of one part in ten thousand for the computation of loop hafnians of matrices with dimension 56, as shown in Fig.~\ref{Fig:error}.

As shown in the previous section, the runtime of the algorithm scales exponentially with the number of photons and linearly with the number of modes. Since the number of photons is the dominant parameter, we benchmark the time taken to calculate the largest event. If $N$ photons are detected at the end of the algorithm, probabilities having at most $N+d$ photons need to be calculated, where $d$ is the cutoff. 

\begin{figure}[t!]
		\centering
\begin{tikzpicture}

\definecolor{color0}{rgb}{0.12156862745098,0.466666666666667,0.705882352941177}

\begin{axis}[
log basis y={10},
tick align=outside,
tick pos=both,
title={Loop hafnian calculation time},
x grid style={white!69.0196078431373!black},
xlabel={Matrix size},
xmin=28.7, xmax=57.3,
xtick style={color=black},
y grid style={white!69.0196078431373!black},
ylabel={Time in seconds},
ymin=0.542564592989256, ymax=46948.8971602024,
ymode=log,
ytick style={color=black}
]
\addplot [semithick, color0, mark=asterisk, mark size=3, mark options={solid}, only marks]
table {%
30 0.909637689590454
32 1.70156979560852
34 3.86939096450806
36 8.75226807594299
38 19.1910259723663
40 44.9143280982971
42 102.492657423019
44 231.156805992126
46 502.338943243027
48 1149.22332143784
50 2499.40737938881
52 5835.70491552353
54 12401.1042561531
56 28003.2474143505
};
\end{axis}

\end{tikzpicture}
		\caption{Computation time of the loop hafnian of a random complex symmetric matrix in long double complex precision using a Standard D96a v4 Microsoft Azure instance with 96 CPUs running at 2.345 GHz. For a matrix of dimension 56, the runtime is approximately seven hours.} \label{Fig:time}
\end{figure}
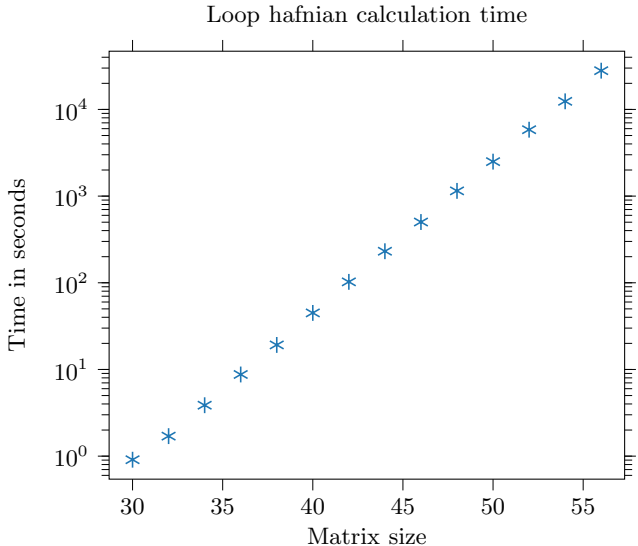

In Fig.~\ref{Fig:time} we benchmark the time it takes to calculate the loop hafnian of a random symmetric complex matrix in long double complex precision up to a total dimension of 56, corresponding to detection of $N=50$ photons with a cutoff of $d=6$. Computations are done on a workstation of 96 CPUs. The matrices are built by sampling each real and imaginary component from the standard uniform distribution, and are then symmetrized. As can be deduced from the figure, the runtime for calculating the loop hafnian of a $56\times 56$ matrix is approximately seven hours. We can estimate the total runtime of the sampling algorithm by approximating the number of times that such loop hafnians need to be computed. 

For a physical system with $m$ modes, an upper bound on the runtime can be obtained by multiplying the runtime of the largest loop hafnian calculation by $m$. However, typically the $N$ photons occur spread out across different modes, so we only calculate the largest loop hafnians for a fraction of the modes. The worst-case occurs when photons are detected in the first $N$ modes, resulting in expensive calculations in the remaining $m-N$ modes. For example, for $m=100$ and $N=50$, we can estimate a runtime of roughly two weeks for generating a single sample on the workstation. 

Finally, the loop hafnian implementation benchmarked here is written for portability and reproducibility. This implies that our implementation of the algorithm is not highly optimized. In the future, we expect to achieve a speed increase of one or two orders of magnitude by low-level optimization of the C++ backend coupled with the use of a modern task-based parallelism library \cite{huang2020cpp} for more efficient load-balancing.

\begin{figure}[!t]
\centering
\begin{tikzpicture}

\definecolor{color0}{rgb}{0.12156862745098,0.466666666666667,0.705882352941177}

\begin{axis}[
log basis y={10},
tick align=outside,
tick pos=both,
title={Loop hafnian relative error},
x grid style={white!69.0196078431373!black},
xlabel={Matrix size},
xmin=28.7, xmax=57.3,
xtick style={color=black},
y grid style={white!69.0196078431373!black},
ylabel={Relative error},
ymin=1.44401039638825e-12, ymax=0.000887531528294776,
ymode=log,
ytick style={color=black}
]
\addplot [semithick, color0, mark=asterisk, mark size=3, mark options={solid}, only marks]
table {%
30 3.62287e-12
32 1.55608e-11
34 5.9249e-11
36 3.43168e-10
38 1.30753e-09
40 2.22745e-09
42 1.21831e-08
44 4.55803e-08
46 1.55652e-07
48 1.29902e-06
50 5.23588e-06
52 1.72386e-05
54 7.40987e-05
56 0.000353754
};
\end{axis}

\end{tikzpicture}
\caption{Relative error in the calculation of the all ones matrix $\mathbbm{1}_n$ for different matrix sizes $n$. The exact value of the loop hafnian for a given size $n$ is given by the $n$-th telephone number $T(n)$~\cite{bjorklund2018faster}; the relative error is given by $ |\text{lhaf}_{\text{numeric}}(\mathbbm{1}_n) -T(n)|/T(n)$. }\label{Fig:error}
\end{figure}
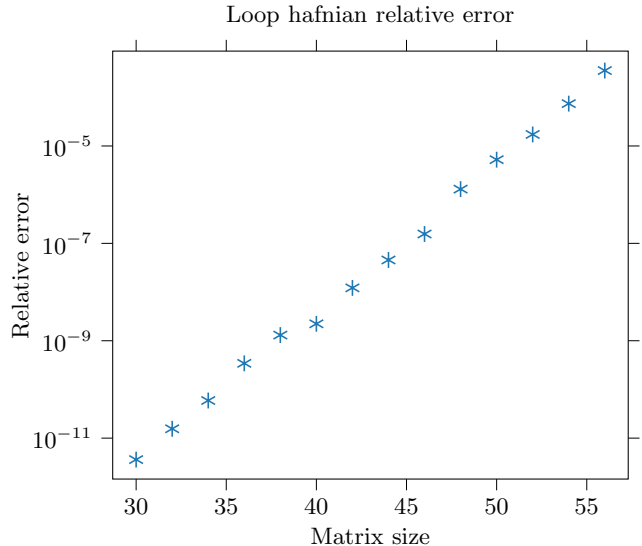

\section{Discussion}\label{sec:discussion}

We have introduced an algorithm for the simulation of Gaussian boson sampling for which the complexity of generating a sample scales, up to polynomial prefactors, like the complexity of calculating a probability amplitude. This results in a quadratic speedup compared to the previous state-of-the-art. The algorithm is exact up to a small error induced by a cutoff dimension, runs in polynomial space, and simulates the most general forms of Gaussian boson sampling.

A remarkable consequence of our result is that Gaussian boson sampling requires roughly twice as many photons than standard boson sampling to reach the same regime of classical simulation. This has potential implications for experimental efforts at demonstrating an advantage over classical simulators. Indeed, for a number of photons $N$ and a number of modes $m$, the complexity of the best algorithm for boson sampling scales as $O(N 2^N)$, whereas our algorithm has runtime $O(m N^3 2^{N/2})$. A possible interpretation is that in Gaussian boson sampling, where photons are generated through squeezing, it is the number of photon pairs that determine complexity; not the number of photons.  \\

\emph{Acknowledgements} -- N.Q. thanks L. Banchi, K. K. Sabapathy and D. Schmid  for fruitful discussions. N.Q. and T.V. thank A. Fumagalli and J. Izaac for technical assistance. The authors thank I. Dhand and O. Di Matteo for comments on the manuscript.
B.A.B. and R.S.C. thank J.F.F. Bulmer, A.E. Jones, S. Paesani, and R.B. Patel for useful discussions. R.S.C acknowledges support from the Engineering and Physical Sciences Research Council (EP/LO15730/1). B.A.B. is supported by a European Commission Marie Skłodowska Curie Individual Fellowship (FrEQuMP, 846073).

\appendix
\onecolumngrid
\section{Proof of correctness}\label{app:correctness}

We prove that the algorithm presented in Sec.~\ref{sec: algorithm} correctly samples from $p(s_1,\ldots,s_m)$. The distribution we are sampling from is given by
\begin{align}
    \tilde{p}(s_1,\ldots,s_m)&=\int d\alpha_2 \cdots d\alpha_m \, p(\alpha_2,\ldots,\alpha_m)p(s_1|\alpha_2,\ldots,\alpha_m)
    p(s_2|s_1,\alpha_3,\ldots,\alpha_m) \cdots p(s_m|s_1,\ldots,s_{m-1}) \nonumber \\
    &=\int d\alpha_2 \cdots d\alpha_m \, p(\alpha_2,\ldots,\alpha_m)
    \frac{p(s_1,\alpha_2,\ldots,\alpha_m)}{p(\alpha_2,\ldots,\alpha_m)}
    \frac{p(s_1,s_2,\alpha_3,\ldots,\alpha_m)}{p(s_1,\alpha_3,\ldots,\alpha_{m})}\cdots
    \frac{p(s_1,\ldots,s_m)}{p(s_1,\ldots,s_{m-1})} \nonumber \\
    &=\int d\alpha_2 \cdots d\alpha_m \,
    \frac{p(s_1,\alpha_2,\ldots,\alpha_m)}{p(s_1,\alpha_3,\ldots,\alpha_m)}
    \frac{p(s_1,s_2,\alpha_3,\ldots,\alpha_m)}{p(s_1,s_2,\alpha_4,\ldots,\alpha_m)}\cdots
    \frac{p(s_1,\ldots,s_{m-1},\alpha_m)}{p(s_1,\ldots,s_{m-1})}
    p(s_1,\ldots,s_m) \nonumber \\
    &=\int d\alpha_2 \cdots d\alpha_m \, p(\alpha_2|s_1,\alpha_3,\ldots,\alpha_m)
    p(\alpha_3|s_1,s_2,\alpha_4,\ldots,\alpha_m)\cdots
    p(\alpha_m|s_1,\ldots,s_m)p(s_1,\ldots,s_m) \nonumber \\
    &=p(s_1,\ldots,s_m),
\end{align}
which is precisely the GBS distribution.
To go from the first to the second line we use the definition of conditional probabilities. From the second to the third line we re-arrange the numerators and denominators of the different fractions. To arrive at the fourth line we use the definition of conditional probability once more. In the last line we use the fact $\int d\alpha_k \, p(\alpha_k|s_1,\ldots,s_{k-1},\alpha_{k+1},\ldots,\alpha_{m})=1$ and integrate in order from $\alpha_2$ to $\alpha_m$.
\twocolumngrid

\bibliography{gbs}

\end{document}